\documentstyle[twoside,fleqn,espcrc2]{article}
\newcommand{\half}{{\scriptstyle{{1\over 2}}}}

\def\beq{\begin{equation}}
\def\eeq{\end{equation}}
\def\bea{\begin{array}}
\def\eea{\end{array}}
\def\beqa{\begin{eqnarray}}
\def\eeqa{\end{eqnarray}}

\def\u1{{U(1)}}
\def\su2{{SU(2)}}
\newcommand{\re}{\relax{\rm I\kern-.18em R}}
\def\cT{{\beta}}

\def\cA{{\cal{A}}}
\def\tr{{\rm tr}} 
\def\Tr{{\rm Tr}} 
\def\pl{{{\cal P}_\infty}}
\def\plo{{{\cal P}_\infty^0}}

\newcommand{\AmS}{{\protect\the\textfont2
  A\kern-.1667em\lower.5ex\hbox{M}\kern-.125emS}}

\title{Exact fermion zero-mode for the new calorons
\vskip-3cm\hfill\small ITEP-TH-27/99; INLO-PUB-13/99\vskip2.6cm
}
\author{M.N. Chernodub\address{Institute of Theoretical and Experimental
Physics,\\B.Cheremushkinskaya 25, Moscow, 117259, Russia},
Thomas C. Kraan\address{Instituut-Lorentz for Theoretical Physics, 
University of Leiden,\\PO Box 9506, NL-2300 RA Leiden, The Netherlands.}
and Pierre van Baal${}^{\rm b}{}$\thanks{Presented by the last author at 
Lattice '99, Pisa, Italy.}}
\begin{document}
\begin{abstract}
We construct the fermion zero-mode for arbitrary charge one $SU(n)$ calorons 
with non-trivial holonomy, both in the finite temperature context 
(anti-periodic boundary conditions in time) and in the Kaluza-Klein 
compactification context (periodic boundary conditions in time). The 
zero-mode is localised on one of the constituent monopoles and we discuss 
a relation to the Callias index theorem.
\end{abstract}
\maketitle
\section{Introduction}
The $SU(n)$ instantons at finite temperature (or calorons) can be seen as
bound states of $n$ constituent monopoles, evident only when the Polyakov 
loop at spatial infinity is non-trivial. In the periodic gauge, 
$A_\mu(t\!+\!\cT,\vec x)\!=\!\!A_\mu(t,\vec x)$, 
\beq
\pl=\lim_{|\vec x|\rightarrow\infty}
P\,\exp(\int_0^\cT A_0(\vec x,t)dt).
\eeq
After a suitable constant gauge transformation, it can be characterised by
$\sum_{m=1}^n\mu_m\!=\!0$ and
\beqa
&&\plo=\exp[2\pi i\,{\rm diag}(\mu_1,\ldots,\mu_n)],\\
&&\mu_1\leq\ldots\leq\mu_n\leq\mu_{n+1}\!\equiv\!1\!+\!\mu_1.\nonumber
\eeqa
Using the classical scale invariance we can always arrange $\beta=1$, as will 
be assumed throughout. A remarkably simple formula for the $SU(n)$ action 
density exists~\cite{PLBN,LAT98}. 
\beqa
&&\Tr F_{\mu\nu}^{\,2}(x)=\partial_\mu^2\partial_\nu^2\log\psi(x),\\
&&\psi(x)=\half\tr(\cA_n\cdots \cA_1)-\cos(2\pi t),\nonumber\\
&&\cA_m\equiv\frac{1}{r_m}\left(\!\!\!\bea{cc}r_m\!\!&|\vec y_m\!\!-\!
\vec y_{m+1}|\\0\!\!&r_{m+1}\eea\!\!\!\right)\left(\!\!\!
\bea{cc}c_m\!\!&s_m\\s_m\!\!&c_m\eea\!\!\!\right),\nonumber
\eeqa
with $r_m\!=\!|\vec x\!-\!\vec y_m|$ the center of mass radius of the 
$m^{\rm th}$ constituent monopole, which can be assigned a mass $8\pi^2\nu_m$,
where $\nu_m\!\equiv\!\mu_{m+1}\!-\!\mu_m$. Furthermore, $c_m\!\equiv\!
\cosh(2\pi\nu_m r_m)$, $s_m\!\equiv\!\sinh(2\pi\nu_m r_m)$, 
$r_{n+1}\!\equiv\! r_1$ and $\vec y_{n+1}\!\equiv\!\vec y_1$.

\section{Monopole constituents}

These generalised caloron solutions can be found~\cite{KrvB} 
using a combination of the Nahm transformation~\cite{Nahm} and the 
Atiyah-Drinfeld-Hitchin-Manin (ADHM) construction~\cite{ADHM}. The latter is 
mainly needed to resolve the delta function singularities that arise in the 
Nahm transformation, although other methods were developed as well~\cite{Lee}.

The Nahm equation for these charge 
one instantons reduces to an abelian problem on the circle, parametrised
by $z$\,mod\,1, 
\beq
\frac{d}{dz}\hat A_j(z)=2\pi i\sum_m\left(y_m^j-y_{m-1}^j\right)\delta(z-\mu_m),
\eeq
giving $\hat A_j(z)=2\pi i y_m^j$, for $z\in[\mu_m,\mu_{m+1}]$. In the monopole
literature $\hat A_j(z)$ is usually denoted by $T_j(z)$. Taking one interval in
isolation, applying the Nahm transformation~\cite{Nahm} gives a single static 
Bogomol'nyi-Prasad-Sommerfeld (BPS) monopole with mass proportional to the 
length ($\nu_m$) of the interval. Taking $|\vec y_n|\!\rightarrow\!\infty$ 
leaves the interval $[\mu_1,\mu_n]$, allowing for the interpretation of an 
$SU(n)$ monopole with $\mu_i$ specifying the eigenvalues of the Higgs field 
at infinity, for which it is crucial they add to zero. Indeed, in the 
periodic gauge $A_0$ tends to a constant at spatial infinity. 

Note that we have to order $\exp(2\pi i\mu_m)$ along the circle to ensure that
the $\nu_i$ add to 1, an ordering inherited by $\mu_m$ when extended to the 
real line by insisting $\mu_{kn+m}\!=\!k\!+\!\mu_m$, for any integer $k$. 
Let us pick one to be labelled by $\mu_1$. All we can guarantee at this point
is that $\sum_{m=1}^n\mu_m=\ell$, an integer. With $\mu_{kn+m}\!=\!k\!+\!\mu_m$,
we find $\sum_{m=1}^n\mu_{m+p}=\ell+p$, for {\em any} integer $p$. A cyclic 
shift of the labels by $p\!=\!-\ell$ proves that there is a {\em unique} choice
of the $\mu_m$ that satisfy eq.~(2). It demonstrates why $\vec y_n$ does play a 
special role, and in the limit $|\vec y_n|\!\rightarrow\!\infty$ one therefore 
has a {\em static} monopole solution~\cite{Nahm}, which can be seen as the 
composite of $n\!-\!1$ BPS monopoles of mass $\nu_m$, located at $\vec y_m$, 
for $m\!=\!1,\cdots,n\!-\!1$. From the general formalism it is clear these 
$n\!-\!1$ monopole constituents are time independent, as was verified 
explicitly for $SU(2)$~\cite{LAT98,KrvB}. Note that our argument demonstrates 
that for $|\vec y_m|\!\rightarrow\!\infty$ with $m\!\neq\!n$, one is left with 
a gauge field that cannot be time independent, even though the resulting 
action density is~\cite{LAT98}.

The significance of one constituent carrying a time dependent field lies in the
fact that the $n$ constituent monopoles form an instanton, and the topological 
charge can be associated to the so-called Taubes-winding~\cite{Taubes}, 
described by a time dependent (gauge) rotation, going full circle when $t$ 
progresses over one period. For $SU(2)$ this can be read-off from the explicit 
expression for the gauge field~\cite{LAT98,KrvB}. We thus conclude that the 
constituent located at $\vec y_n$ is the one that carries this Taubes-winding, 
even though its action density is time independent for well-separated 
constituents. This conclusion can also be drawn from the formalism developed 
in ref.~\cite{Lee}, see also ref.~\cite{SUSY}.
 
\section{Fermion zero-mode}

The basic ingredient in the construction of caloron solutions is a Greens 
function defined by 
\beq
\left(\!\!D_z^2\!+\!r^2(x;z)\!+\!\!\sum_m\!\delta_m(z)\!\right)\!\!
\hat f_x(z,z')\!=\!\delta(z\!-\!z'),
\eeq
where $D_z\!=\!(2\pi i)^{-1}\!\partial_z\!-\!t$, $r^2(x;z)\!=\!r_m^2(x)$ for 
$z\in[\mu_m,\mu_{m+1}]$ and $\delta_m(z)=\delta(z\!-\!\mu_m)|\vec y_m\!-\!
\vec y_{m-1}|/2\pi$. A similarity with the impurity scattering
problem allows for a straightforward solution~\cite{PLBN}, which we present
here for the case that $\mu_m\!\leq\!z'\!\leq\!z\!\leq\!\mu_{m+1}$ (extended to
$z<z'$ by $\hat f_x(z',z)=\hat f_x^*(z,z')$)
\beqa
&\hskip-3mm\hat f_z(z,z')\!=\!\!\frac{\pi e^{2\pi i t(z\!-\!z')}}{r_m\psi}\Bigl(
e^{-2\pi it}\sinh\left(2\pi(z-z')r_m\right)\!+\hskip-4.1mm\nonumber\\
&<\!v_m(z')|\cA_{m\!-\!1}\cdots \cA_1\cA_n\cdots \cA_m|w_m(z)\!>\!\Bigr),
\eeqa
where the spinors $v_m$ and $w_m$ are defined by
\beqa
&v_m^1(z)=-w_m^2(z)=\sinh\left(2\pi(z\!-\!\mu_m)r_m\right),\nonumber\\
&v_m^2(z)=\hphantom{-}w_m^1(z)=\cosh\left(2\pi(z\!-\!\mu_m)r_m\right).
\eeqa
For the zero-mode densities we find
\beq
|\Psi_z(x)|^2=-(2\pi)^{-2}\partial_\mu^2\hat f_x(z,z),
\eeq
derived exactly as for $SU(2)$~\cite{ZM2}, not repeated here. 
With the gauge field in the periodic gauge one has, $\Psi_z(t\!+\!1,\vec x)=
\exp(2\pi i z)\Psi_z(t,\vec x)$. To obtain the finite temperature fermion 
zero-mode one puts $z=\half$, whereas for the fermion zero-mode with periodic 
boundary conditions, relevant in supersymmetric applications, one takes $z=0$.

In figure 1 we show a typical $SU(3)$ caloron, illustrating that also for
$n>2$ the fermion zero-modes are localised on one of the constituents. 
\begin{figure}[htb]
\vspace{4.5cm}
\includegraphics{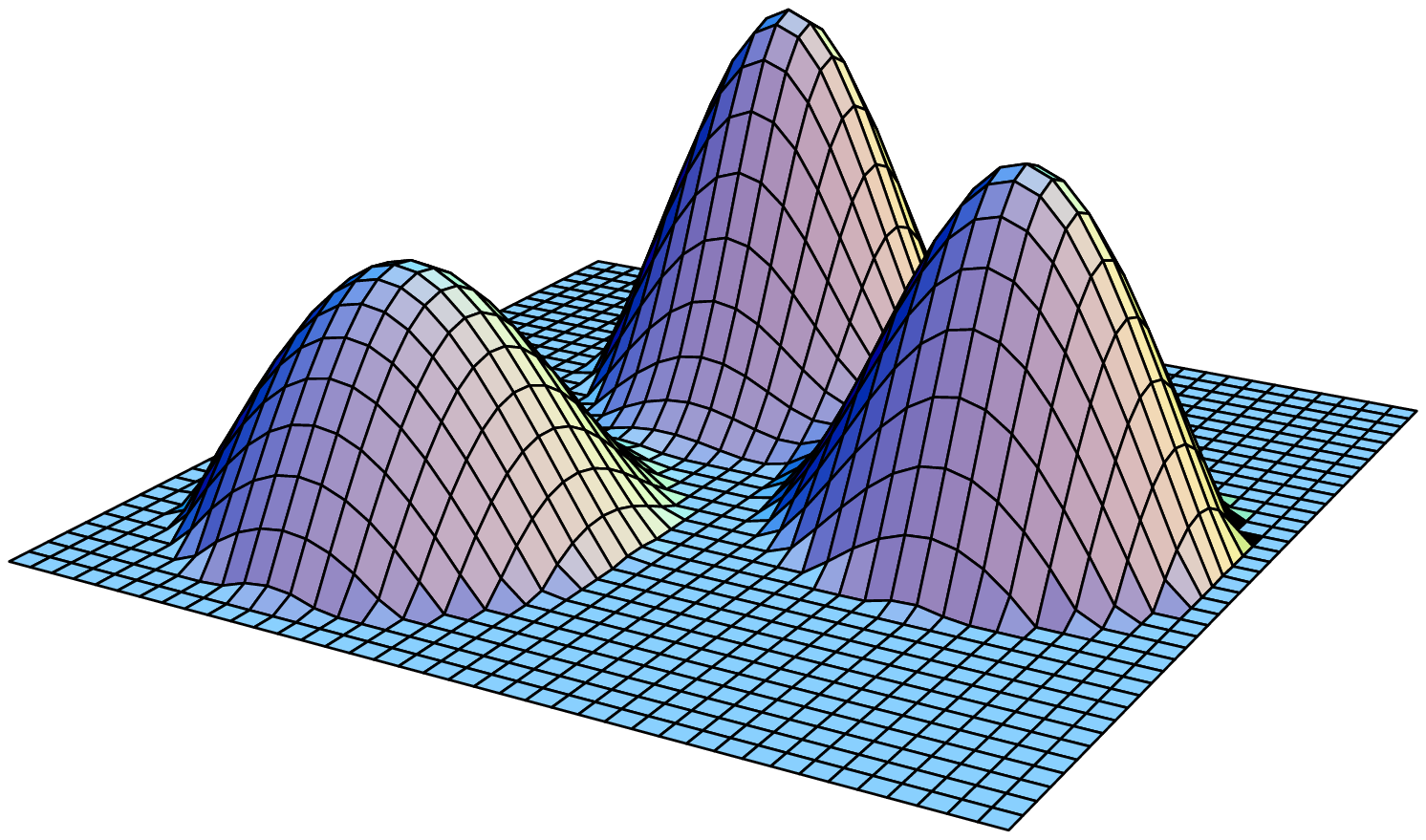}
\includegraphics{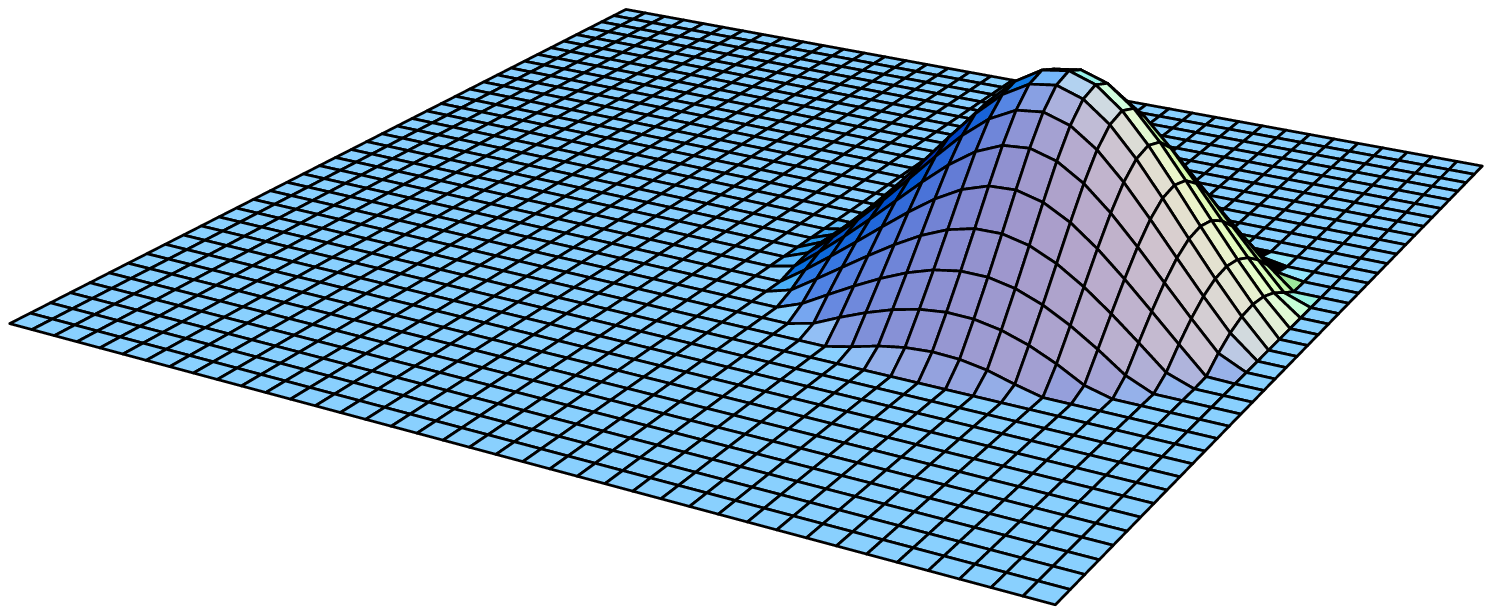}
\includegraphics{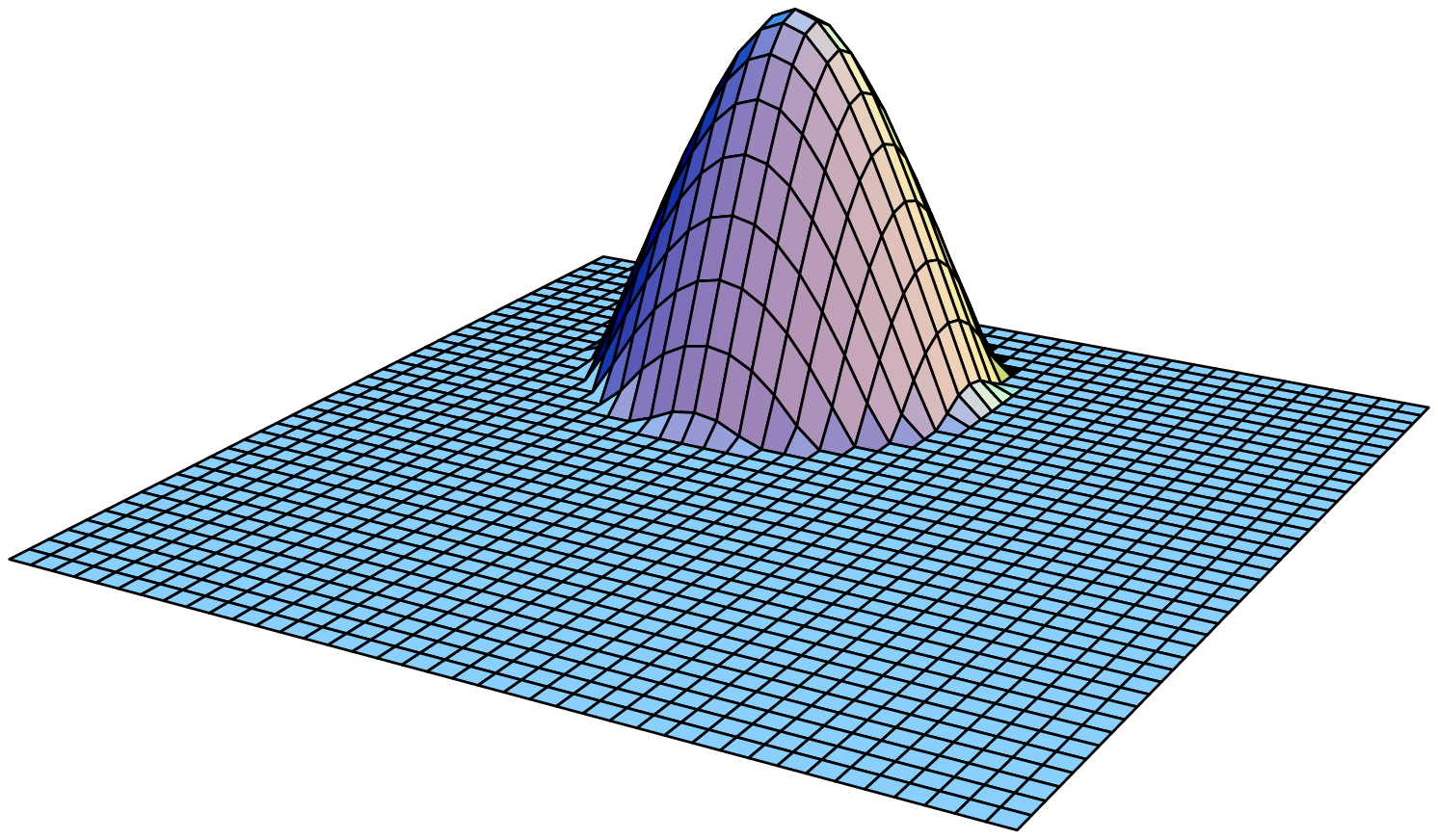}
\caption{The action densities (top) for the $SU(3)$ caloron, cut off 
at $1/(2e)$, on a logarithmic scale, with $(\mu_1,\mu_2,\mu_3)=(-17,-2,19)/60$ 
for $t\!=\!0$ in the plane defined by $\vec y_1\!=\!(-2,-2,0)$, $\vec y_2=
(0,2,0)$ and $\vec y_3=(2,-1,0)$, for $\cT=1$, with masses $8\pi^2\nu_i$,
$(\nu_1,\nu_2,\nu_3)=(0.25,0.35,0.4)$.  On the bottom-left is shown 
the zero-mode density for fermions with anti-periodic boundary conditions 
in time and on the bottom-right for periodic boundary conditions, at equal 
logarithmic scales, cut off below $1/e^5$.}
\end{figure}
This localisation can be established easily in the limit of large
$|\vec y_i\!\!-\!\vec y_{i+1}|$ for all $i$,
in which case one finds, when $z\in[\mu_m,\mu_{m+1}]$,
\beqa
\!\hat f_x(z,z)\!\!=\!\!\frac{\sinh[2\pi(z\!-\!\mu_m)r_m]\sinh[2\pi(\mu_{m+1}\!
-\!z)r_m]}{r_m\sinh[2\pi\nu_mr_m]/2\pi}\hskip-8mm\nonumber
\eeqa
making explicit that the location of the zero-mode is determined by the 
interval that contains the appropriate value of $z$. From eq.~(2) it follows 
that $\mu_1\!\leq\!0\!\leq\!\mu_n$, such that the periodic zero-mode is 
associated to the {\em static} constituent at $\vec y_m$, with $\mu_m\!\leq\!0
\!\leq\!\mu_{m\!+\!1}$. This is precisely the condition for the existence 
of a zero-mode given by the Callias index theorem~\cite{Call} (see 
also the appendix of ref.~\cite{JdB}). Due to the static background (for 
well-separated constituents), time dependence of the zero-mode would be 
of the form $\exp(2\pi ikt)$ for $k$ integer, shifting $z\!=\!0$ by $k$,
out of the interval that allows for a zero-mode. 

Allowing for $k\!=\!\pm\half$, for which $\exp(2\pi ikt)$ turns the periodic 
zero-mode anti-periodic, we can have situations where this anti-periodic 
zero-mode is associated to one of the static monopole constituents. A specific 
example for $SU(3)$ where this occurs is $(\mu_1,\mu_2,\mu_3)\!=\!(-0.48,-0.03, 
0.51)$, yielding $(\nu_1,\nu_2,\nu_3)\!=\!(0.45,0.54,0.01)$. {\em Both} the 
periodic and the anti-periodic zero-mode are associated to the $2^{\rm nd}$ 
constituent. We note that, apart from the fact that the $3^{\rm rd}$ 
constituent is nearly massless, both zero-modes are very broad since 
min$(z\!-\!\mu_2,\mu_3\!-\!z)\!=\!0.03$ for $z\!=\!0$ and $0.01$ for 
$z\!=\!\half$. For $SU(2)$ $z\!=\!0$ is always midway between 
$\mu_1$ and $\mu_2$ and $z\!=\!\half$ midway between $\mu_2$ and 
$\mu_3\!=\!1\!+\!\mu_1$). When $z$ coincides with $\mu_i$, the
zero-mode is no longer normalisable, which is the origin of the delta 
function singularities in the Nahm transformation.

\section{Conclusions}

In conclusion, for well-separated constituents the fermion zero-mode is 
localised to a single constituent. For $SU(2)$ the anti-periodic zero-mode
is always associated to the constituent that carries Taubes-winding~\cite{ZM2}.
For $SU(n\!>\!2)$ this is also typically true (see fig.~1), in particular 
when well localised, something that may be significant for developing a model 
for the QCD vacuum that combines monopoles and instantons~\cite{LAT98,KrvB,ZM2}.
However, exceptions exist where both the periodic and anti-periodic zero-mode 
are associated to (possibly the same) static constituent(s), although this 
tends to be accompanied by nearly massless constituents, and rather
delocalised zero-modes. 

\section*{Acknowledgements}
We thank, Jan de Boer for pointing us to the connection with the 
Callias theorem, Tam\'as Kov\'acs and Mikhail Voloshin for discussions, 
Diego Bellisai for correspondence and Margarita Garc\'{\i}a P\'erez for 
collaboration in the early stages. MNCh thanks the Theoretical Physics 
Institute at the University of Minnesota and the Institute-Lorentz for 
hospitality. This work was supported in part by INTAS grant RFBR 95-0681. 
MNCh was supported by the INTAS fellowship YSF 98-95 and TCK by a grant 
from FOM/SWON. PvB thanks the Aspen Center of Physics, where part of 
this work was done, and workshop organisers Urs Heller and Rajamani 
Narayanan, and parti\-cipants, for creating a stimulating atmosphere.

\end{document}